\documentclass[final,conference]{IEEEtran} 

\usepackage{graphicx}
\usepackage{amsmath,amsfonts,amssymb}
\usepackage{citesort}
\usepackage{subfigure}
\usepackage{url}
\usepackage{color}

\newtheorem{theorem}{Theorem}

\newtheorem{lemma}{Lemma}
\newtheorem{corollary}{Corollary}[theorem]

\newcommand{\thh}[2]{\theta(\hat{#1},\hat{#2})}

\newcommand{\br}{{\bf r}}
\graphicspath{{.}}

\begin{document}


\title{Uncertainty Principles for Signal Concentrations}%
\author{%
        \authorblockN{Ram Somaraju}
        \authorblockA{Australian National University and National ICT Australia\\
        {\tt Ram.Somaraju@rsise.anu.edu.au}}
        \and%
        \authorblockN{Leif W. Hanlen}
        \authorblockA{National ICT Australia  and Australian National University\\
        {\tt Leif.Hanlen@nicta.com.au}}%
        }%
        
\maketitle

\begin{abstract}
Uncertainty Principles for concentration of signals into truncated subspaces are considered. The ``classic'' Uncertainty Principle is explored as a special case of a more general operator framework. The time-bandwidth concentration problem is shown as a similar special case. A spatial concentration of radio signals example is provided, and it is shown that an Uncertainty Principle exists for concentration of single-frequency signals for regions in space. We show that the uncertainty is related to the volumes of the spatial regions.
\end{abstract}

\section{Introduction}\label{introduction}

Signal processing, at its most fundamental level, concerns the extraction of information (signal) from a system under various constraints. Typically, communication engineers consider the signal as being embedded in noise, and although the signal may have spatial, temporal or bandwidth constraints, they are usually concerned with noise as the limiting factor. In the following work, we will consider a fundamental limit to signal extraction  without impediments due to noise. We shall pose the following question:

\begin{itshape}
\noindent Given a signal (wavefield) which has energy in one volume $V_A$, how well can we constraint that signal to have energy in \emph{another} volume $V_B$? 
\end{itshape}

This question motivates examination of \emph{Uncertainty Principles} (UP's). We may regard uncertainty as a basic limit on signal information extraction, without requiring a model for noise -- since any noise can only further hinder our efforts.
Uncertainty Principles  have gained great popularity since  Heisenberg. The famous example is Heisenberg's Uncertainty Principle: it is impossible to exactly measure the location \emph{and} momentum of a particle simultaneously. This is a special case of a more general framework, which may be applied (via various Hilbert space techniques)  to a range of scenarios. 

In a communication theory setting, a similar UP has been well known: that a signal cannot be arbitrarily confined in both time and frequency. The reader is directed to \cite{Slep:1976} for a discussion. The work of \cite{Pollak0161,Pollak0161b,Pollak0162} has formalised this result, although without explicit reference to the operator theoretic nature of the problem\footnote{Landau provides some work in this regard~\cite{Landau65}.}. The time-bandwidth concept of a temporal signal provides us with a great deal of intuition  and we shall draw heavily on the classic works. 

There is a significant distinction between \emph{essential dimensionality} results such as~\cite{Kennedy:0203b,Slep:1976} and \emph{uncertainty} results~\cite{Selig01,FaiVarISIT00}. Dimensionality results (eg. $2WT$) may be seen as a counting of the number of degrees of freedom a signal (or function) has within a particular set of constraints -- much like the rank of a matrix. Uncertainty tells us to what extent  a signal will achieve all constraints, and how uniquely that solution is specified by the constraints.
 %

The remainder of this paper is arranged as follows: In Section~\ref{classup} we collate classic results in uncertainty and formulate UP's in terms of Operator Theoretic forms. Section~\ref{comms} develops an Uncertainty Principle for communication between volumes with a particular form of operator channel. We provide a free-space example. Conclusions are drawn, and proofs are consigned to the appendix.

\newcommand{\hil}{\mathcal{H}}
\newcommand{\lt}{$L^2[\mathbb{R}]$}
\newcommand{\ltt}{L^2[\mathbb{R}]}
\newcommand{\cd}{\mathcal{D}}
\newcommand{\cb}{\mathcal{B}}
\newcommand{\ltn}[1]{\|#1\|_2}
\newcommand{\io}{\int_{-\Omega}^{\Omega}}
\newcommand{\itt}{\int_{-T/2}^{T/2}}
\newcommand{\sino}[1]{\frac{\sin{\Omega(#1)}}{\pi(#1)}}

\section{Uncertainty Principles: a review}\label{classup}
Before explaining the Uncertainty Principle, we develop some
relevant notation.
\subsection{Notation} Following Selig~\cite{Selig01}, let $\hil$ be a Hilbert
space with inner product $\langle\cdot,\cdot\rangle $ and norm
$\|\cdot\| \equiv \langle\cdot,\cdot\rangle ^{1/2}$. Further, let
$A$  be a linear operator with domain $\mathcal{D}(A) \subseteq
\hil$ and range in $\hil$. We then define the normalised
expectation value, $\tau_A(f)$ and standard deviation
(uncertainty), $\sigma_A(f)$ of the operator $A$ with respect to
$f\in\mathcal{D}(A)$ to be~\cite{Selig01}
\begin{align}
  \tau_A(f) &\equiv \frac{\langle Af,f\rangle }{\langle f,f\rangle } \\
  \sigma_A(f) &\equiv \|(A-\tau_A(f))f\|.
\end{align}
The adjoint of operator $A$, $A^\dagger$ is defined using the
following equation~\cite{kre78}
\begin{equation}\label{adjdef}
    \langle Ax,y\rangle  = \langle x,A^\dagger y\rangle  \textrm{\ \ }\forall x\in \mathcal{D}(A),y \in
    \mathcal{D}(A^\dagger).
\end{equation}
Furthermore, $A$ is said to be Hermitian or self-adjoint if $A =
A^\dagger$. For a self-adjoint operator
\begin{equation}\label{seladj}
    \langle Ax,y\rangle  = \langle x,Ay\rangle  \textrm{\ \ }\forall x,y \in
    \mathcal{D}(A).
\end{equation}
Any operator that obeys equation~\ref{seladj} is said to be
symmetric.

Given two linear operators $A$ and $B$ with domains
$\mathcal{D}(A)$ and $\mathcal{D}(B)$ respectively, the commutator
is defined as
\begin{equation}\label{comm}
    [A,B] \equiv AB - BA
\end{equation}
and the anti-commutator is defined as
\begin{equation}\label{comm}
    [A,B]_+ \equiv AB + BA
\end{equation}
with domains $\mathcal{D}(AB)\cap\mathcal{D}(BA)$ for either one.
Operators $A$ and $B$ are said to \emph{commute} with each other
if $[A,B] = 0$. Otherwise they are called non-commutative
operators.

Also, let $L^2[\mathbb{R}]$ be the set of all square integrable
functions defined on the real line with norm $\|f\|_2 = \langle
f,f\rangle^{1/2}$. Here the inner product
$\langle\cdot,\cdot\rangle$ is defined as
\begin{equation}\label{l2inner}
    \langle f,g\rangle = \int_{-\infty}^\infty
    f(t)\overline{g(t)}dt.
\end{equation}
Consider $\ltn{f}^2$ as the energy of the function
$f\in\ltt$. The angle between two non-zero functions $f$ and $g$
is defined as
\begin{equation}\label{theta}
    \theta(f,g) = \cos^{-1}\frac{\Re\{\langle
    f,g\rangle\}}{\ltn{f}\ltn{g}}.
\end{equation}
We also define
\begin{equation}\label{ft}
    \hat{f}(\omega) = \int_{-\infty}^{\infty} f(x) e^{-i\omega
    x}dx
\end{equation}
to be the Fourier transform of $f(x)$ whenever this integral
exists.

\subsection{Classic Uncertainty Principle}
The classical uncertainty principle states that the values of two
non-commuting observers such as position and momentum cannot be
precisely determined in any quantum state. That is, the standard
deviation of two non-commuting operators cannot be made
arbitrarily small simultaneously. The following theorem is a
general statement of this phenomenon.

\begin{theorem}[theorem 3.4,\cite{Selig01}] \label{th:upsym} If $A$ and $B$ are two
symmetric operators on a Hilbert space $\hil$, then
\setlength{\arraycolsep}{0.0em}
\begin{multline}\label{uprinc}
    \left\| (A-a)f\right\|\left\| (B-b)f\right\| \geq \frac{1}{2}\left\{\left|\langle [A,B]f,f\rangle\right|^2 \right.\\
    +\left.\left|\langle[A-aI,B-bI]_+f,f\rangle\right|^2\right\}^{1/2}
\end{multline}
for all $f \in \mathcal{D}(AB)\cap\mathcal{D}(BA)$ and all $a,b
\in \mathbb{R}$. Equality holds precisely when $(A-a)f$ and
$(B-b)f$ are scalar multiples of one another.
\end{theorem}

A special case of this theorem bounds the standard deviation of
two non-commuting operators as explained in the following
corollary which we get by substituting $a = \sigma_A(f)$ and $b =
\sigma_B(f)$ in theorem~\ref{th:upsym}.

\begin{corollary}\label{hunc}
If $A$ and $B$ are two symmetric operators on a Hilbert space
$\hil$, then \setlength{\arraycolsep}{0.0em}
\begin{equation}\label{uprinc}
    \sigma_A(f)\sigma_B(f) \geq \frac{1}{2}\sqrt{|\langle [A,B]f,f\rangle|^2}
\end{equation}
 for all $f \in
\mathcal{D}(AB)\cap\mathcal{D}(BA)$ and all $a,b \in \mathbb{R}$.
Equality holds precisely when $(A-\tau_A(f))f$ and
$(B-\tau_B(f))f$ are scalar multiples of one another.
\end{corollary}

A special case of this corollary is the following Heisenberg
Uncertainty Principle which states that a function and its Fourier
transform cannot be arbitrarily localised.
\begin{theorem}[theorem 6.1,~\cite{Selig01}]\label{th:uptf}
Let $f \in L^2(\mathbb{R}), \|f\| = 1$ and set
\begin{align*}
  x_o &= \int x|f(x)|^2 dx 
  &\omega_o = \int \omega|\hat{f}(\omega)|^2 d\omega \\
  \Delta x &= \int (x-x_o)^2|f(x)|^2 dx \quad
  &\Delta\omega = \int (\omega-\omega_o)^2|\hat{f}(\omega)|^2 d\omega 
\end{align*}
whenever these integrals exist. Then 
\begin{equation*}
\Delta x\Delta \omega \geq
\pi/2,
\end{equation*}
 where equality is attained iff $$f(x) =
(r/\pi)^{1/4}e^{i\omega_o x}e^{-r(x-x_o)^2/2}$$ for any $r>0$.
\end{theorem}

Theorem~\ref{th:uptf} gives valuable insight into how localized a function
can be in both time and frequency. If one defines $\Delta x$ and
$\Delta \omega$ to be the measure of approximate time duration and
bandwidth of the signal respectively, then Theorem~\ref{th:uptf} says that
the product of time duration and bandwidth of a function is
bounded from below by $\pi/2$: if the time-spread gets
very small, the frequency-spread must be large and vice-versa. 

Though a good qualitative tool, this is inadequate
for the purposes of signal processing. Theorem~\ref{th:uptf} does not
answer the question: given a bandlimited function (i.e.
$\hat{f}(\omega) = 0$ for $\omega \notin [-\Omega,\Omega]$) how
much of the energy of $f$ is `concentrated' in finite duration of
time. This would be useful in answering the question: ``given a
bandlimited channel, how much of the transmitted signal can a
receiver measure over a finite period of time?''

\subsection{Uncertainty principle for energy concentrations}
\label{tmfreq} Landau and Pollak~\cite{Pollak0161b} proposed that for the purposes of
signal processing, a more relevant uncertainty principle should
use sharper measures of concentrations in time and frequency than
the ones used in Heisenberg's principle~\cite{Pollak0161b}. To help
derive their uncertainty principle~\cite{Pollak0161b} define $\cd = \{f:
f\in \ltt, f(t) = 0 \ \forall |t| > T/2 \}$ to be the class of all
time-limited functions and $\cb = \{f:f\in \ltt, \hat{f}(\omega) =
0 \ \forall |\omega| > \Omega\}$ to be the class of all band
limited functions. Here, $\hat{f}(\omega)$ is the Fourier
transform of $f(t)$ as defined in equation~\ref{ft}. Also $T$, the
time duration and $W = \Omega/2\pi$, the bandwidth are fixed for
the remainder of this paper. It is easy to prove~\cite{Pollak0161} that
$\cd$ and $\cb$ are complete subspaces of \lt.

Define the projection operators
\newline $B:\ltt\to\cb$ and $D:\ltt\to\cd$ as follows
\begin{align}
  D f(t) &= \left\{\begin{array}{cc}
    f(t), & |t| \leq T/2 \\
    0, & |t| > T/2 \\
  \end{array}\right. \\
  B f(t) &= \frac{1}{2\pi}\int_{-\Omega}^{\Omega}
  \hat{f}(\omega)e^{j\omega t} d\omega
\end{align}

Using these operators we can calculate the fraction of energy,
$\alpha^2$ of any function $f\in\ltt$ in the finite duration of
time $[-T/2,T/2]$.
\begin{equation}\label{fractime}
    \alpha^2 = \frac{\ltn{Df}^2}{\ltn{f}^2}.
\end{equation}
Similarly, we can calculate, $\beta^2$, the fraction of energy of
a function in a finite bandwidth $[-\Omega,\Omega]$.
\begin{equation}\label{fracbw}
    \beta^2 = \frac{\ltn{Bf}^2}{\ltn{f}^2}.
\end{equation}
\cite{Pollak0161b} show that $\alpha$ and $\beta$ cannot be arbitrarily
large. Specifically, they prove
\begin{theorem}[theorem 2,~\cite{Pollak0161b}]
Let $0\leq\alpha,\beta\leq 1$. Then there exists a function
$f\in\ltt$, $\ltn{f} =1$ with $\ltn{Df} = \alpha$ and $\ltn{Bf} =
\beta$ if and only if $(\alpha,\beta) \neq (0,1) \textrm{ or }
(1,0)$ and
\begin{equation*}
 \cos^{-1}\alpha + \cos^{-1}\beta \geq
    \cos^{-1}\sqrt{\lambda_0}
\end{equation*}
where, $\lambda_0$ is the largest eigenvalue of the equation
\begin{equation*}
    \lambda \psi = BD \psi
\end{equation*}
\end{theorem}
This theorem constrains the possible values of $\alpha$ and
$\beta$ because $\lambda_0 < 1$~\cite{Pollak0161}. So any function cannot have arbitrarily large fractions
of energy in both a finite-time duration and a finite-frequency
bandwidth. We will show in the next section that this theorem is a
special case of a more general theorem just like the Heisenberg
Principle is a special case of the classical uncertainty principle
as alluded to by~\cite{Pollak0161b}. We show that this more general
theory can be used to understand communication through arbitrary
channels.


\newcommand{\ar}[1]{a_{R_#1}(\mathbf{r}_R)}
\newcommand{\ard}[1]{a_{R_#1}^\dagger(\mathbf{r}_R)}
\section{Communication between finite volumes and the uncertainty principle}
\label{comms} The derivations above are for a
function defined on the real line and its Fourier transform, however the
principle can be extended to arbitrary transforms defined
on $\mathbb{R}^n$. 
The key property is that the subspaces
$\cb$ and $\cd$ form nonzero minimum angles. Before deriving these
results, we explain the physical model.

\begin{figure}
\begin{center}\setlength{\unitlength}{0.85mm}
\begin{picture}(80,40)
\put(5,0){\includegraphics[width=0.3\textwidth]{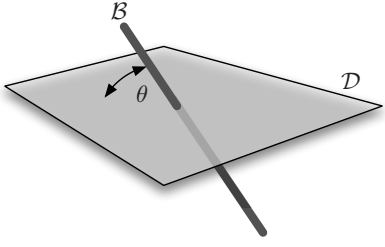}}
\put(28,22){\small $\theta$}
\put(24,35){\small $\mathcal{B}$}
\put(60,24){\small $\mathcal{D}$}
\end{picture}
\caption{Subspaces $\mathcal{B}$ and $\mathcal{D}$, with  minimum angle $\theta$. Point of intersection is $0$ vector. Here, $\mathcal{B},\mathcal{D}\subset\mathbb{R}^3$, below, $\mathcal{B},\mathcal{D}\subset\mathcal{H}$ with dimension $\aleph_0$.}
\end{center}
\end{figure}

\subsection{Physical Problem and Notation}
Consider communication using scalar waves between two volumes
$V_T\textrm{, } V_R\subset\mathbb{R}^3$, where $\mathbb{R}^3$ is
the standard three dimensional Euclidian space. Let $L^2(V)$ be
the space of all square integrable functions defined on volume
$V\subset\mathbb{R}^3$ with the standard inner product. Following~\cite{mil00},  assume there are sources\footnote{This
is the time independent wave function. The complete time-dependant
wave function, for exponential time dependence may be written as
$F(\br,t) = f(\br)\exp\{i\omega t\}$. However the theory can
easily be extended by defining volumes $V\subset\mathbb{R}^4$
where the fourth coordinate is time.} $f\in L^2(\mathbb{R}^3)$
that generate waves $\hat{f}\in L^2(\mathbb{R}^3)$ governed by
some transformation $\Gamma:L^2(\mathbb{R}^3)\to
L^2(\mathbb{R}^3)$:
\begin{equation}\label{trans}
    \hat{f} = \Gamma f.
\end{equation}
Here, $L^2[\mathbb{R}^3]$ is the set of all square integrable
functions defined on $\mathbb{R}^3$ with the standard inner
product. If we only have sources in the transmitting volume and
can only measure waves within the receiving volume we will need to
consider restrictions of $f$ and $\hat{f}$ to $V_T$ and $V_R$
respectively. Let $\chi_V$ be the characteristic (truncation) operator:
\begin{equation}
\chi_V f(\br) = %
\begin{cases}
    f(\br) &\br \in V, \\
    0 &\textrm{otherwise.}
\end{cases}
\end{equation}
We say that a function $f$ is $\epsilon$-concentrated in volume
$V$ in some norm $\|\cdot\|$ if
\begin{equation*}
    \frac{\|\chi_V f\|^2}{\|f\|^2} \geq 1-\epsilon^2.
\end{equation*}

With this notation in place we can now look at the set of
functions $\cd$ that a receiver can measure and the set of
functions $\cb$ a transmitter can generate:
\begin{align}
  \cd &= \{f: f = \chi_{V_R} g, g\in L^2[\mathbb{R}^3] \} \\
  \cb &= \{f: f = \Gamma \chi_{V_T} g, g\in L^2[\mathbb{R}^3]\}.
\end{align}
If the spaces $\cd$ and $\cb$ are complete, we can define
projection operators $D:L^2[\mathbb{R}^3]\to\cd = \chi_{V_R}$ and
$B:L^2[\mathbb{R}^3]\to\cb$. Note that $\cb \neq \Gamma
\chi_{V_T}$. We can also define an angle between these two
subspaces as follows:
\begin{equation}\label{angsub}
    \theta(\cb,\cd) = \inf_{\begin{array}{c}\scriptstyle f\in\cb,g\in\cd\\\scriptstyle f\neq 0,g\neq
    0\end{array}} \theta(f,g).
\end{equation}

\subsection{Uncertainty Principle for arbitrary subspaces}
We can again think of $\ltn{f}^2$ as the energy of a function.
Then $\alpha^2 = \ltn{Df}^2/\ltn{f}^2$ is the fraction of energy
of $f$ in the space of receiver functions and $\beta^2 =
\ltn{Bf}^2/\ltn{f}^2$ is the fraction of the energy of $f$ in the
space of transmitter functions. We can now prove the uncertainty
principle that constrains the possible range of values that
$\alpha$ and $\beta$ can take provided the subspaces $\cb$ and
$\cd$ form a non-zero minimum angle.
\begin{theorem}\label{th:uptheta} Let subspaces $\cb$ and $\cd$
form a non-zero minimum angle $\theta_0$ then
\begin{equation}\label{uprinctheta}
    \cos^{-1} \alpha + \cos^{-1} \beta \geq \theta_0.
\end{equation}
\end{theorem}

This theorem has a very simple physical interpretation. If the
space of all the functions that a transmitter can generate and the
space of all the functions a receiver can receive form a non-zero
minimum angle then there exist no functions that can have
arbitrarily large fractions of energy in these two spaces of
functions. We can calculate this minimum angle by calculating the
norm of the operator $\|BD\|_2$ and this is the subject of our
next theorem.
\begin{theorem}\label{th:minimum angle}
The angle between two complete subspaces $\cb$ and $\cd$ with
projection operators $B$ and $D$ is
\begin{equation}\label{minangle}
    \theta(\cb,\cd) = \cos^{-1}\ltn{BD}
\end{equation}
\end{theorem}

\subsection{General Uncertainty principle} We prove a slightly modified
version of the general uncertainty theorem proved in~\cite{don89}.
In the following $L^1[V]$, $L^2[V]$ and $L^\infty[V]$ are spaces
of functions defined on $V\subset\mathbb{R}^3$ with finite $L_1$
($\|\cdot\|_1 = \int_V |\cdot|$), $L_2$ ($\|\cdot\|_2 = \int_V
|\cdot|^2$) and $L_\infty$ ($\|\cdot\|_\infty = \sup_{V} |\cdot|$)
norms.

\begin{theorem}[Generalized Uncertainty]\label{unccom}
Suppose $\Gamma(f) = \hat{f}$ and $f\in L^1[\mathbb{R}^3]\bigcap\
L^2[\mathbb{R}^3$ and $\hat{f}\in L^2[\mathbb{R}^3]\bigcap\
L^\infty[\mathbb{R}^3]$ and satisfies
\begin{enumerate}
    \item $\ltn{f} = \alpha\ltn{\hat{f}}$
    \item $\|\hat{f}\|_\infty \leq \beta\|f\|_1.$
\end{enumerate}
Also, suppose $f$ is $\epsilon_T$-concentrated to $V_T$ in the
$L_1$ norm and $\hat{f}$ is $\epsilon_R$-concentrated to $V_R$ in
the $L_2$ norm. Then,
\begin{equation*}
    |V_T||V_R|\alpha^2\beta^2 \geq
    {(1-\epsilon_T)^2(1-\epsilon_R^2)}
\end{equation*}
\end{theorem}

The theorem has a very simple physical interpretation for
communication between finite volumes. Firstly, by requiring
$\ltn{f} = \alpha\ltn{\hat{f}}$, we ensure that the energy of the
received signal is proportional to that of the transmitting
signal. So $\alpha$ determines the attenuation in the signal and
we expect it to be greater than 1. Secondly, $\|\hat{f}\|_\infty
\leq \beta\|f\|_1$ can be thought of as a stability condition
(i.e. bounded input gives bounded output).
As a direct corollary to Theorem~\ref{unccom} we obtain an uncertainty principle for communicating volumes
\begin{corollary}[Volumetric Communication Uncertainty]{\it
Let the transmitting volume $V_T$ be finite, and the conditions for Theorem~\ref{unccom} hold. Then 
\begin{equation*}
    |V_T||V_R|\alpha^2\beta^2 \geq
    {(1-\epsilon_R^2)}
\end{equation*}
Since $f$ must be
perfectly concentrated in $V_T$ and so $\epsilon_T = 0$. ie. the maximum fraction of energy that can be inside the
receiving volume is bounded from above. }
\end{corollary}


\section{Conclusions}
We investigated Uncertainty Principles for the transmission of waves between volumes, using a generalised version of the classic Heisenberg UP. We have provided a framework for future UP developments where volumetric constraints on signals induce a limit to signal resolution. Application of operator-theoretic tools has provided a means to develop concent

\footnotetext{National ICT Australia is funded
through the Australian Government's \emph{Backing Australia's Ability initiative}, in part through the Australian Research Council.}

\appendix


\begin{proof}[outline theorem~\ref{th:upsym}] Let
$A,B,a,b$ and $f$ be as stated in the theorem. We summarize~\cite{Selig01}.
From the
Cauchy-Schwarz inequality we have \setlength{\arraycolsep}{0.0em}
\begin{multline}\label{thm}
  2\|(B-b)f\|\|(A-a)f\| =2[\Im\{\langle(B-b)f,(A-a)f\rangle\}^2\\
                         +\Re\{\langle(B-b)f,(A-a)f\rangle\}^2]^{1/2}
\end{multline}
 $\Re\{z\}$ and
$\Im\{z\}$ denote real and imaginary parts of
$z$. 
\begin{equation}\label{impart}
  2i\Im\{\langle(B-b)f,(A-a)f\rangle\} = \langle[A,B]f,f\rangle 
\end{equation}
Using symmetry of the operators  and the fact that scalar multiplication
commutes with all linear operators. Similarly
\begin{equation} \label{repart}
  2\Re\langle(B-b)f,(A-a)f\rangle = \langle[A-aI,B-bI]_+f,f\rangle
\end{equation}
Re-arranging~\eqref{impart}, \eqref{repart}, \eqref{thm} completes the proof.
\end{proof}


\begin{proof}[outline  Theorem~\ref{th:uptf}]
The theorem is proved trivially by noting that $\sigma_A(f) =
\Delta x$ and $\sigma_B(f) = \Delta\omega/2\pi$~\cite{Selig01};
where, $Af = \cdot f$ and $Bf = -if'$~\cite{Selig01}.
\end{proof}


\begin{lemma}\label{le:sumang}
Let $f,g,h \in L^2[\mathbb{R}^3]$. Then,
\begin{equation}\label{sumang}
    \theta(f,g) \leq \theta(f,h) + \theta(g,h).
\end{equation}
\end{lemma}

\begin{proof}[Proof of Lemma~\ref{le:sumang}] Let $\hat{f} = f/\|f\|_2$. Then
\begin{align}
 \theta(f,g) &= \cos^{-1} \frac{\Re\{\langle\ltn{f}\hat{f},\ltn{g}\hat{g}\rangle\}}{\ltn{f}\ltn{g}} 
  = \cos^{-1}\Re\{\langle\hat{f},\hat{g}\rangle\} \nonumber\\
  &= \theta(\hat{f},\hat{g})\label{blah}
\end{align}
Let $\theta(f,g) \neq 0$. Otherwise, there is nothing to prove.
Also, let $S = span\{f,g\}$ be the space of all functions spanned
by $f$ and $g$. Then this space is complete and we can write $h$
as~\cite{kre78}
\begin{equation*}
   \hat{h} = h^\| + h^\bot,
\end{equation*}
where, $h^\| \in S$ and $h^\bot$ is orthogonal to both $f$ and
$g$. Because $\|h^\|\|_2\leq \|\hat{h}\|_2 = 1$ we have
\begin{align}
 \theta(f,h) &= \thh{f}{h} 
  = \cos^{-1}\Re\{\langle\hat{f},\hat{h}\rangle\} 
  = \cos^{-1}\Re\{\langle\hat{f},h^\|\rangle\} \nonumber\\
  &\geq \cos^{-1}\Re\left\{\frac{\langle \hat{f},h^\|\rangle}{\ltn{h^\|}}\right\} \nonumber\\
  &= \theta(\hat{f},h^\|)\label{theta1}
\end{align}
Similarly,
\begin{equation}\label{theta2}
   \theta(g,h) \geq \theta(\hat{g},h^\|)
\end{equation}
Now, if $\theta(\hat{g},h^\|) = 0$ the proof is trivial
as $$\theta(f,g) =
\theta(\hat{f},\hat{g})=\theta(\hat{f},h^\|)\leq\theta(f,h)\leq\theta(f,h)+\theta(g,h).$$

If $\theta(\hat{g},h^\|) \neq 0$, let
\begin{align}
 \hat{h}_1 &= h^\|/\|h^\|\|_2, \\
 \hat{h}_2 &= \frac{g-\hat{h}_1\langle
g,\hat{h}_1\rangle}{\|(g-\hat{h}_1\langle g,\hat{h}_1\rangle\|_2}
\end{align}
be two unit vectors that are orthogonal to each other and whose
span is $S$. We can therefore write,
\begin{align}
 \hat{f} &= a_1\hat{h}_1 + a_2 \hat{h}_2 \\
 \hat{g} &= b_1\hat{h}_1 + b_2 \hat{h}_2
\end{align}
Where, $a_1=a_1'+ ia_1'',a_2=a_2'+ ia_2'',b_1=b_1'+ib_1''$ and
$b_2 = b_2'+ib_2''$ are complex numbers. From the orthogonality of
$\hat{h}_1$ and $\hat{h}_2$ we have
\begin{align}
 \cos\theta(\hat{f},h^\|) &= a_1' \label{th2}\\
 \cos\theta(\hat{g},h^\|) &= b_1' \label{th1}\\
 \cos\theta(\hat{f},\hat{g}) &= \Re\{a_1^*b_1 + a_2^*b_2\}\\
 &= a_1'b_1' + a_2'b_2' + a_1''b_1'' + a_2''b_2''
\end{align}
 $a_i^*$ is the complex conjugate of $a_i$. From
orthogonality of $\hat{h}_1$ and $\hat{h}_2$ and 
$\hat{f}$ and $\hat{g}$ have unit norm
\begin{align}
 a_1'^2 + a_1''^2 + a_2'^2 + a_2''^2 &= 1 \label{suma}\\
 b_1'^2 + b_1''^2 + b_2'^2 + b_2''^2 &= 1 \label{sumb}
\end{align}
Consider $(a_1'',a_2',a_2'')$ and $(b_1'',b_2',b_2'')$ as
two three dimensional vectors, the Cauchy-Schwarz inequality
gives 
\begin{equation*}
 \sqrt{a_1''^2+a_2''^2+a_2'^2}\sqrt{b_1''^2+b_2''^2+b_2'^2} \geq |a_1''b_1''+a_2''b_2''+a_2'b_2'|
\end{equation*}
Here, $|.|$ denotes the absolute value of a real number.
\begin{equation}\label{agb}
   a_1''b_1''+a_2''b_2''+a_2'b_2' \geq
   -\sqrt{a_1''^2+a_2''^2+a_2'^2}\sqrt{b_1''^2+b_2''^2+b_2'^2}
\end{equation}
\begin{align}
 &\cos(\theta(\hat{f},h^\|)+\theta(\hat{g},h^\|)) \notag\\
 &=\cos(\theta(\hat{f},h^\|))\cos(\theta(\hat{g},h^\|))
  {-}\sin(\theta(\hat{f},h^\|))\sin(\theta(\hat{g},h^\|)) \notag\\
  &= a_1'b_1' - \sqrt{1-a_1'^2}\sqrt{1-b_1'^2} \label{step1} \\
  &= a_1'b_1' - \sqrt{a_1''^2+a_2''^2+a_2'^2}\sqrt{b_1''^2+b_2''^2+b_2'^2}\label{step2}\\
  &\leq a_1'b_1' + a_1''b_1''+a_2''b_2''+a_2'b_2' \label{step3}\\
  &= \cos(\theta(\hat{f},\hat{g}))\notag
\end{align}
\setlength{\arraycolsep}{5pt} We get equation~\eqref{step1} from
equations~\eqref{th2} and~\eqref{th1}. Equations~\eqref{suma}
and~\eqref{sumb} are used to get~\eqref{step2} and finally we use
inequality~\eqref{agb} to get~\eqref{step3}. Now, from the
monotonicity of $\cos$, we have
\begin{equation}\label{sjsj}
   \theta(\hat{f},\hat{g}) \leq
   \theta(\hat{f},h^\|)+\theta(\hat{g},h^\|).
\end{equation}
Substituting inequalities~\eqref{theta1} and~\eqref{theta2} into
the above and using equation~\eqref{blah} proves the lemma.
\end{proof}


\begin{proof}[Proof outline Theorem~\ref{th:uptheta}] From the definition of $\theta(f,g)$ 
\begin{align*}
  \cos(\theta(f,Df)) &= \frac{\Re\{\langle f,Df\rangle\}}{\ltn{Df}\ltn{f}} 
   =  \frac{\Re\{\langle Df,Df\rangle\}}{\ltn{Df}\|f\|_2}
   = \frac{\ltn{Df}}{\ltn{f}} \\
   &= \alpha.
\end{align*}
Noting %
$f = Df + f - Df$ and $\langle f-Df, Df\rangle = 0$. Similarly $\beta = \cos(\theta(f,Bf))$. In order to complete our proof
we use Lemma~\ref{le:sumang}. Therefore
\begin{align}
  \cos^{-1}\alpha + \cos^{-1}\beta &= \theta(f,Df) + \theta(f,Bf) \\
         &\geq \theta(Df,Bf)\\
         &\geq \theta_0\label{p4:3}
\end{align}
where \eqref{p4:3} is from  $Df\in\cd$ and $Bf \in\cb$
and these two subspaces have the minimum angle $\theta_0$.
\end{proof}

\begin{proof}[Proof of Theorem~\ref{th:minimum angle}]
The angle between two subspaces is
\begin{align}
      \theta(\cb,\cd) &= \inf_{\begin{array}{c}\scriptstyle f\in\cb,g\in\cd\\\scriptstyle f\neq 0,g\neq
    0\end{array}} \theta(f,g)
\end{align}
Using $\cos\theta(f,g)$ from the proof of the last theorem, we
can write
\begin{align}
  \cos \theta(\cb,\cd) &= \sup_{\begin{array}{c}\scriptstyle f\in\cb,g\in\cd\\ \scriptstyle \ltn{f}= 1,\ltn{g}=1\end{array}} \Re\{\langle f,g \rangle\} \\
   &= \sup_{\begin{array}{c}\scriptstyle f\in L^2[\mathbb{R}^3],g\in L^2[\mathbb{R}^3]\\ \scriptstyle \ltn{f}= 1,\ltn{g}=1\end{array}} \Re\{\langle Bf,Dg \rangle\} \label{p5:2}\\
   &= \sup_{\begin{array}{c}\scriptstyle f\in L^2[\mathbb{R}^3],g\in L^2[\mathbb{R}^3]\\ \scriptstyle \ltn{f}= 1,\ltn{g}=1\end{array}} \Re\{\langle f,BDg \rangle\}\label{p5:3}\\
   &= \|BD\|_2
\end{align}
where \eqref{p5:2} is from self-adjointness of
$B$ and \eqref{p5:3} from the definition of the operator norm.
\end{proof}

\begin{proof}[Proof of Theorem~\ref{unccom}]
\begin{align*}
    \|f\|_2^2 &=\alpha^2\|\hat{f}\|_2^2 \\
            &\leq\alpha^2(1-\epsilon_R^2)^{-1}|V_R|\|\hat{f}\|_\infty^2\\
            &\leq\alpha^2(1-\epsilon_R^2)^{-1}|V_R|\beta^2\|f\|_1^2\\
            &\leq\alpha^2(1-\epsilon_R^2)^{-1}|V_R|\beta^2(1-\epsilon_T)^{-2}\int_{V_T}|f|\\
            &\leq\alpha^2(1-\epsilon_R^2)^{-1}|V_R|\beta^2(1-\epsilon_T)^{-2}|V_T|\|f\|_2^2
\end{align*}
We get the last step using the Cauchy-Schwarz inequality. Rearranging the above inequality gives the required result.
\end{proof}


\end{document}